\documentclass[runningheads]{llncs}
\usepackage[T1]{fontenc}
\usepackage{amsmath}
\usepackage{algorithm}
\usepackage{algpseudocode}

\usepackage{amsmath,amssymb}
\usepackage[T1]{fontenc}
\usepackage{graphicx}
\begin{document}
\title{Minimally Supervised Hierarchical Domain Intent Learning for CRS}
\author{Safikureshi Mondal\inst{1}\orcidID{0000-0001-8163-6905} \and
Subhasis Dasgupta\inst{2}\orcidID{0000-0002-0754-0515} \and
Amarnath Gupta\inst{3}\orcidID{0000-0003-0897-120X}
}

\authorrunning{S. Mondal et al.}

\institute{San Diego Supercomputer Center, University of California San Diego, USA \\
\email{\{s4mondal\inst{1}, sudasgupta\inst{2}, a1gupta\inst{3}\}@ucsd.edu}
}

\maketitle              % typeset the header of the contribution
\begin{abstract}

Modeling domain intent within an evolving domain structure presents a significant challenge for domain-specific conversational recommendation systems (CRS). The conventional approach involves training an intent model using utterance-intent pairs. However, as new intents and patterns emerge, the model must be continuously updated while preserving existing relationships and maintaining efficient retrieval. This process leads to substantial growth in utterance-intent pairs, making manual labeling increasingly costly and impractical. In this paper, we propose an efficient solution for constructing a dynamic hierarchical structure that minimizes the number of user utterances required to achieve adequate domain knowledge coverage. To this end, we introduce a neural network-based attention-driven hierarchical clustering algorithm designed to optimize intent grouping using minimal data. The proposed method builds upon and integrates concepts from two existing flat clustering algorithms—DEC and NAM—both of which utilize neural attention mechanisms.

We apply our approach to a curated subset of 44,000 questions from the business food domain. Experimental results demonstrate that constructing the hierarchy using a stratified sampling strategy significantly reduces the number of questions needed to represent the evolving intent structure. Our findings indicate that this approach enables efficient coverage of dynamic domain knowledge without frequent retraining, thereby enhancing scalability and adaptability in domain-specific CSRs.

\end{abstract}

\section{Introduction}

\begin{figure}[htbp]
  \centering
  \includegraphics[width=.80\textwidth]{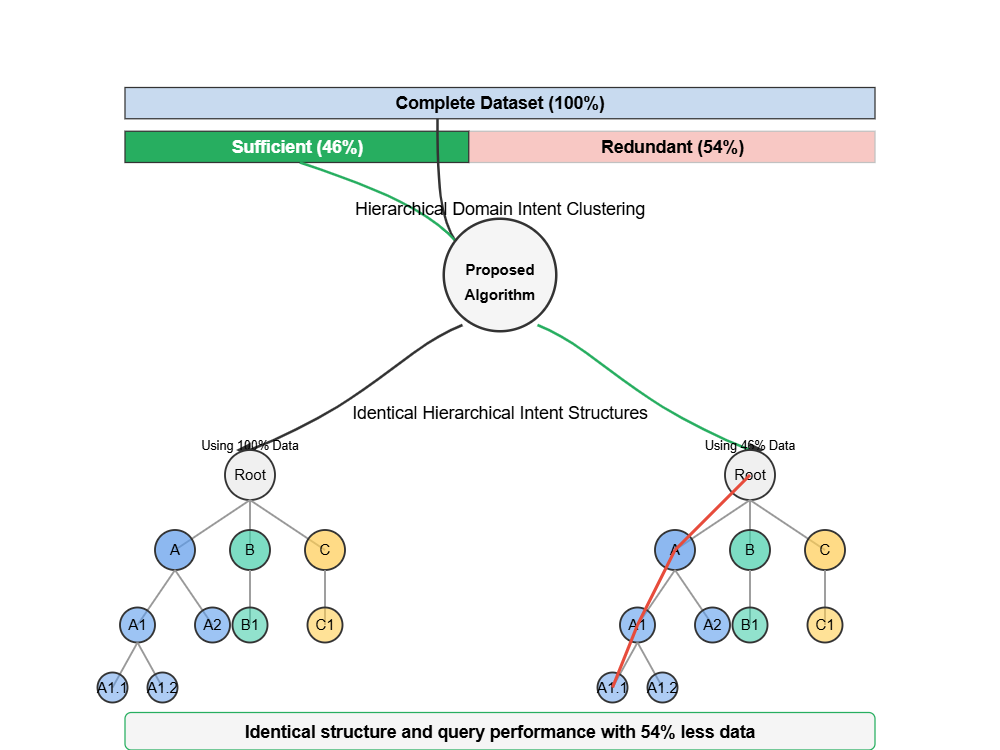} % Adjust width/path
  \caption{Minimal Data Requirement for Optimal Hierarchical Intent Clustering}
  \label{fig:fig6}
\end{figure}

Recent advances in conversational recommendation systems (CRS) have highlighted the critical challenge of maintaining accurate domain intent representations as knowledge structures evolve~\cite{wu2023iterative}. It requires continuous incremental changes to evolving user needs and business requirements, creating inherent challenges for intent discovery and representation~\cite{zhang2020discovering}. Traditional intent modeling approaches relying on static utterance-intent pair datasets face significant scalability limitations as systems encounter novel intents and linguistic patterns in dynamic business structures~\cite {liu2021open}~\cite{haponchyk2018supervised}. This paper addresses the fundamental tension between the construction of an efficient hierarchical domain intent structure methodology with an efficient sampling strategy for finding the domain coverage and minimizing manual annotation efforts through optimized clustering strategies that balance computational efficiency with semantic preservation~\cite{patent1}~\cite{barnabo2023supervised}. The primary challenge lies in proposing a domain intent tree structure to optimize domain knowledge coverage while minimizing the required number of user utterances by an efficient sampling strategy. This optimization is crucial for developing dynamic domain intent systems that can efficiently handle new questions without constantly updating the underlying knowledge structure. Our research addresses this challenge by investigating clustering techniques that can identify representative utterances covering the full intent space. Figure~\ref{fig:fig6} demonstrates how our hierarchical domain intent clustering algorithm efficiently creates identical intent structures using only 46\% of the original dataset. By identifying and removing redundant data (54\%), we achieve the same quality hierarchical organization and query performance while significantly reducing computational requirements and lowering resource barriers for implementation.
Clustering techniques play a crucial role in organizing user utterances into coherent groups that represent underlying intents. Existing clustering-based methods for intent recognition often suffer from inefficiencies in data utilization. Existing methods, such as flat clustering techniques like Deep Embedded Clustering (DEC)~\cite{guo2018deep} and Neural Attention Models (NAM)~\cite{he2017unsupervised}, struggle to preserve hierarchical relationships between intents, leading to fragmented domain structures. Traditional approaches either rely on exhaustive datasets or fail to optimize cluster coverage effectively, leading to increased annotation costs and suboptimal performance in dynamic domains. However, the challenge lies in achieving optimal cluster coverage with minimal user input, which is essential for reducing annotation requirements and enhancing the system's ability to learn dynamically. Moreover, many current systems are not designed to handle the hierarchical nature of intents or adapt to evolving domain knowledge structures. This work aims to address these limitations by proposing a novel framework that optimizes cluster coverage using efficient sampling strategies.\\
We propose a methodology that combines hierarchical clustering with strategic sampling techniques to identify the minimum set of utterances needed to represent the complete intent space. The hierarchical clustering algorithm is built with integration of neural network-based attention mechanism such as Deep Embedded Clustering (DEC)~\cite{guo2018deep} and Neural Attentive Models (NAM)~\cite{he2017unsupervised} flat clustering concepts, and we establish a framework for efficient domain intent management that reduces annotation requirements while maintaining comprehensive coverage. Using a subset of a dataset of 44,000 business domain questions, our experiments determine the optimal number of representative utterances needed to cover the full intent space. This approach provides a foundation for self-learning dynamic domain knowledge structures that can efficiently process new questions without requiring continuous updates.\\
In summary, the primary contribution of the work is as follows:
\begin{enumerate}
    \item A minimally-supervised framework that combines efficient sampling with hierarchical clustering for efficient dynamic domain knowledge structure.
    \item Determination of adequate domain coverage for constructing self-learning domain knowledge structure.
    \item Study the performances of our proposed hierarchical clustering algorithm.
\end{enumerate}

\section{Related Work}
In recent years, intent classification and clustering methods have been used for conversational recommendation systems to handle the domain intents. However, intent clustering has primarily focused on supervised approaches that require substantial labeled data. Liu and Ian R. Lane~\cite{liu2016attention} introduced attention-based RNN models for joint intent detection and slot filling, and in~\cite{chen2019bert}, the author explored BERT-based architectures for intent classification. These approaches, while effective, depend heavily on large labeled datasets. These are actively supervised approaches to intent analysis.\\
The unsupervised and semi-supervised approaches have emerged as alternatives to annotation requirements. The author, Guo, introduced Deep Embedded Clustering (DEC)~\cite{guo2018deep}, which uses deep neural networks to learn feature representations and cluster assignments simultaneously. Lin proposed clustering user utterances to discover new intents automatically~\cite{lin2019discovering}.  Meanwhile, author Lee developed a Neural Attentive Model (NAM) for intent induction that leverages attention mechanisms.~\cite{he2017unsupervised}.
Traditional density-based approaches like ITER-DBSCAN~\cite{chatterjee2020intent} show annotation requirements through iterative refinement but require complete dataset reprocessing for each domain update.\\
Managing evolving domain knowledge has been explored in several contexts. In \cite{madotto2020continual} research paper, the author proposed methods for continual learning in task-oriented dialogue systems, while in~\cite{kalimuthu2019incremental}, the author introduced an approach for incremental domain adaptation. These works highlight the importance of efficient knowledge update mechanisms but often overlook the optimization of the utterance set required for comprehensive coverage.\\
Our research builds upon these foundations while addressing the specific challenge to build the hierarchical systems to optimize domain knowledge coverage with minimal utterances. It provides a novel approach to enabling efficient self-learning in domain-specific conversational recommendation systems.

\section{Proposed Methodology}
In conversational recommendation systems (CRS), domain intent structures evolve dynamically as users introduce new queries that require multi-granular intent recognition. The intent discovery is needed for hierarchical clustering due to some advantages for handling domain intent like preserving the hierarchical relationship (e.g.,{restaurants $\rightarrow$ cuisine type $\rightarrow$ dietary constraints}), adapting or updating new intent without retraining on full utterance pairs, and minimizing labeling costs by reusing existing hierarchical clusters as templates for new intents.\par Our proposed algorithm, hierarchical clustering, is based on Neural Attention and Adaptive Entropy Merging, and it dynamically maintains evolving domain intent structures by optimizing hierarchical clusters of utterances. It operates in four phases. An algorithmic data-flow diagram is shown to understand the proposed methodology in Figure.~\ref{fig:fig1}. The four phases of algorithm are: 

\begin{figure}[htbp]
  \centering
  \includegraphics[width=0.9\textwidth]{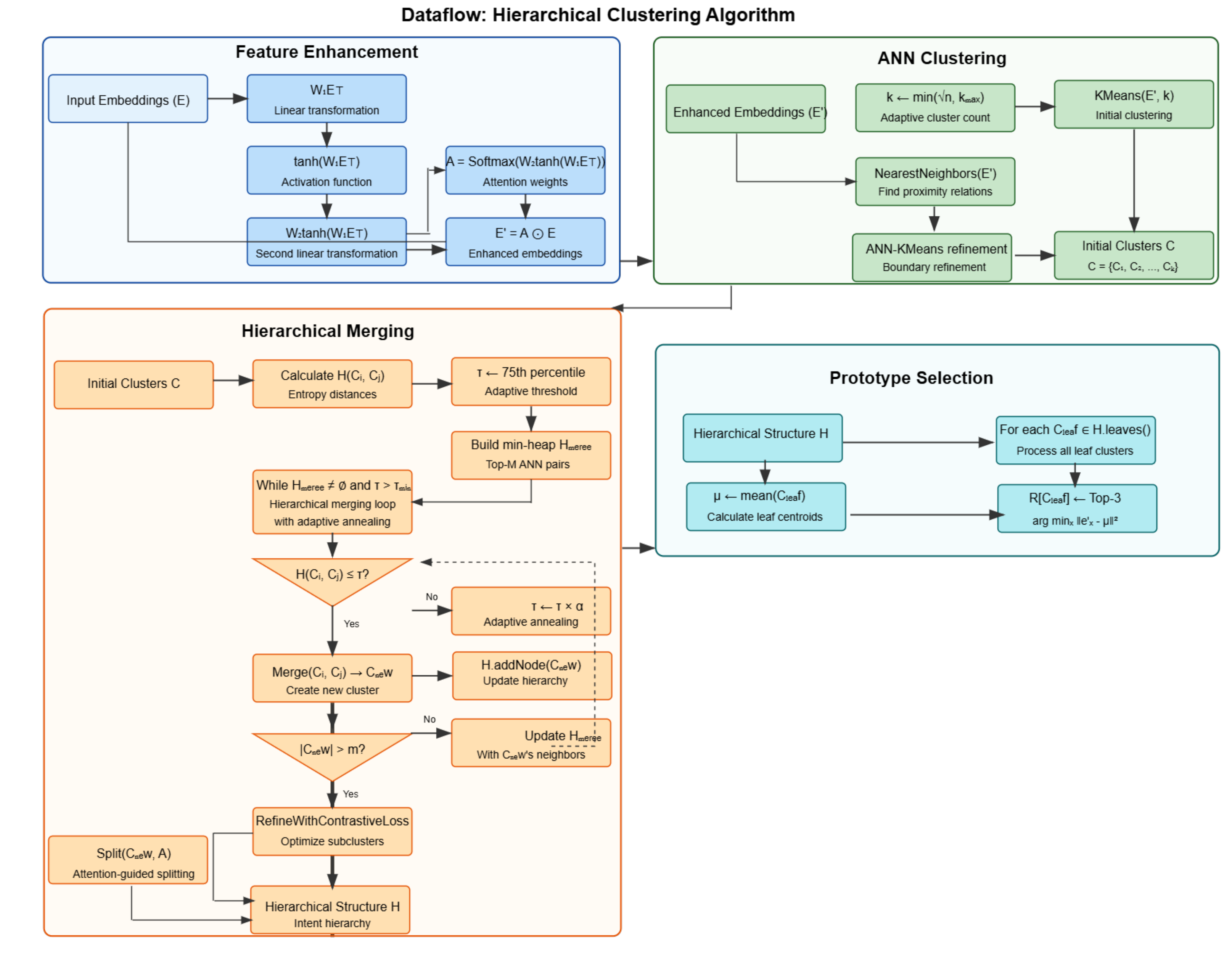} % Adjust width/path
  \caption{Dataflow of Hierarchical clustering algorithm}
  \label{fig:fig1}
\end{figure}

\begin{enumerate}
    \item Feature Enhancement with Neural Attention
    \item Initial Clustering using ANN
    \item Entropy-Guided Hierarchical Merging
    \item Representative Prototype Selection
\end{enumerate} The proposed algorithm are depicted in \ref{slb1} section.
In phase 1, attention-based feature refinement is used  
using taring with pre-train with reconstruction loss and fine-tuning with DEC loss. By using adaptive cluster count, building the ANN index, and assigning utterances to clusters using ANN search for centroid updates, phase 2 generates the initial clusters. Iteratively merges clusters using attention-weighted entropy to preserve semantic coherence, leveraging ANN-accelerated pairwise comparisons and adaptive threshold annealing to build a multi-granular hierarchy (e.g., grouping "Italian restaurants"$\rightarrow$ "vegan" vs. "gluten-free"). Here, contrastive refinement subroutine procedure is used for sub-cluster refinement. Phase 4 selects minimal, centroid-aligned utterances as prototypes for each leaf cluster, enabling efficient intent generalization and low-cost updates.\\
Our hierarchical clustering algorithm strategically integrates key aspects from both DEC and NAM to create a more effective intent clustering system. From DEC, we adopt the concept of learning optimized embedding representations during the clustering process, but modify it to work within a hierarchical framework. Specifically, while traditional DEC jointly optimizes embeddings and flat cluster assignments using a target distribution, our approach incorporates this embedding enhancement within an entropy-based hierarchical merging framework.
The attention mechanism from NAM is adapted in our feature enhancement phase, where we train a specialized AttentionModule to weight the importance of different dimensions in the embedding space. Unlike standard NAM implementations focused on sequence data, our attention module operates directly on the embedding vectors, learning to emphasize features that are most discriminative for hierarchical cluster formation. This attention-weighted transformation creates enhanced embeddings better suited for capturing the hierarchical relationships between intents.
By combining DEC's embedding optimization with NAM's selective feature weighting, our algorithm achieves a more natural hierarchical structure that better reflects the inherent relationships between different intent categories, while requiring substantially less training data than approaches using either concept in isolation.

\subsection{Proposed Hierarchical Algorithm}\label{slb1}

\begin{algorithm}
\caption{Efficient Hierarchical Clustering via Neural Attention and Adaptive Entropy Merging}
\label{alg:ago1}
\begin{enumerate}
\item \textbf{Input:} User utterances $\mathbf{X} = \{x_1, \dots, x_n\}$, attention dim $h$, ANN neighbors $M$, threshold $\tau_{\text{min}}$
\item \textbf{Output:} Hierarchical intent structure $\mathcal{H}$, prototypes $\mathcal{R}$
\item \textbf{Feature Enhancement:}
  \begin{enumerate}
  \item $\mathbf{E} \gets \text{}Embedding(\mathbf{X})$ \quad\% Initial embeddings
  \item $\mathbf{A} \gets \text{Softmax}(\mathbf{W}_2 \tanh(\mathbf{W}_1 \mathbf{E}^\top))$ \quad\% Attention weights
  \item $\mathbf{E}' \gets \mathbf{A} \odot \mathbf{E}$ \quad\% Refined embeddings
  \end{enumerate}
\item \textbf{ANN Clustering:}
  \begin{enumerate}
  \item $k \gets \min(\sqrt{n}, k_{\text{max}})$
  \item $\mathcal{C} \gets \text{ANN-kmeans}(\mathbf{E}', k)$ \quad\% Adaptive initialization
  \end{enumerate}
\item \textbf{Hierarchical Merging:}
  \begin{enumerate}
  \item $\tau \gets 75^{\text{th}}$ percentile of $\{\mathcal{H}(C_i, C_j)\}$
  \item Build min-heap $\mathcal{H}_{\text{merge}}$ with top-$M$ ANN pairs
  \item \textbf{While} $\mathcal{H}_{\text{merge}} \neq \emptyset$ \textbf{and} $\tau > \tau_{\text{min}}$:
    \begin{enumerate}
    \item $(C_i, C_j) \gets \mathcal{H}_{\text{merge}}.\text{pop}()$
    \item \textbf{If} $\mathcal{H}(C_i, C_j) > \tau$:
      \begin{enumerate}
      \item $\tau \gets \tau \times \alpha$ \quad\% Adaptive annealing
      \item \textbf{continue}
      \end{enumerate}
    \item $C_{\text{new}} \gets \text{Merge}(C_i, C_j)$
    \item $\mathcal{H}.\text{addNode}(C_{\text{new}})$ \quad\% Update hierarchy
    \item \textbf{If} $|C_{\text{new}}| > m$: \quad\% Contrastive refinement
      \begin{enumerate}
      \item $\text{subclusters} \gets \text{Split}(C_{\text{new}}, \mathbf{A})$
      \item $\text{RefineWithContrastiveLoss}(\text{subclusters}, \tau_{\text{contrast}})$
      \end{enumerate}
    \item Update $\mathcal{H}_{\text{merge}}$ with $C_{\text{new}}$'s ANN neighbors
    \end{enumerate}
  \end{enumerate}
\item \textbf{Prototype Selection:}
  \begin{enumerate}
  \item \textbf{For each} $C_{\text{leaf}} \in \mathcal{H}.\text{leaves}()$:
    \begin{enumerate}
    \item $\boldsymbol{\mu} \gets \frac{1}{|C_{\text{leaf}}|} \sum_{x \in C_{\text{leaf}}} \mathbf{e}'_x$
    \item $\mathcal{R}[C_{\text{leaf}}] \gets \text{Top-3 } \arg\min_{x} \|\mathbf{e}'_x - \boldsymbol{\mu}\|_2$
    \end{enumerate}
  \end{enumerate}
\end{enumerate}
\end{algorithm}

The Efficient Hierarchical Clustering via Neural Attention and Adaptive Entropy Merging algorithm presents a novel approach to organizing user utterances into a hierarchical intent structure. 
In the first phase, we enhance the representation of user utterances using transformer-based embeddings and neural attention. \begin{itemize}
\item Initially, we obtain embeddings $\mathbf{E}$ for all utterances $\mathbf{X} = {x_1, \dots, x_n}$ using a transformer model:
\begin{equation}
\mathbf{E} = \text{Embedding}(\mathbf{X})
\end{equation}
\item We then apply an attention mechanism to focus on the most discriminative features. The attention weights $\mathbf{A}$ are computed using a two-layer neural network with a $\tanh$ activation function:
\begin{equation}
    \mathbf{A} = \text{Softmax}(\mathbf{W}_2 \tanh(\mathbf{W}_1 \mathbf{E}^\top))
\end{equation}
where $\mathbf{W}_1 \in \mathbb{R}^{h \times d}$ and $\mathbf{W}_2 \in \mathbb{R}^{1 \times h}$ are learnable parameters, with $h$ representing the attention dimension and $d$ the embedding dimension.

\item Finally, we refine the embeddings by applying the attention weights using element-wise multiplication:
\begin{equation}
    \mathbf{E}' = \mathbf{A} \odot \mathbf{E}
\end{equation}
This produces attention-enhanced embeddings $\mathbf{E}'$ that emphasize the most relevant semantic information.
\end{itemize}
The second phase involves initial clustering using Approximate Nearest Neighbors (ANN):
\begin{itemize}
\item We adaptively determine the number of clusters $k$ based on the dataset size:
\begin{equation}
k = \min(\sqrt{n}, k_{\text{max}})
\end{equation}
where $n$ is the number of utterances and $k_{\text{max}}$ is a predefined maximum.
\item We then apply an ANN-accelerated k-means algorithm to the enhanced embeddings:
\begin{equation}
    \mathcal{C} = \text{ANN-kmeans}(\mathbf{E}', k)
\end{equation}
This approach provides an efficient initialization for the subsequent hierarchical structure by leveraging approximate nearest neighbor search to speed up standard k-means.\end{itemize}
The third phase constitutes the core of our algorithm, where we build a hierarchical structure through adaptive merging:
\begin{itemize}
\item We initialize the merging threshold $\tau$ as the 75th percentile of all pairwise cluster similarities:
\begin{equation}
\tau = \text{75th percentile of } {\mathcal{H}(C_i, C_j)}
\end{equation}
where $\mathcal{H}(C_i, C_j)$ represents the similarity between clusters $C_i$ and $C_j$. 
\item We construct a min-heap $\mathcal{H}_{\text{merge}}$ containing the top-$M$ ANN pairs of clusters based on their similarities.

\item The merging process continues while the heap is not empty and the threshold exceeds the minimum threshold $\tau_{\text{min}}$:

\begin{itemize}
    \item Extract the most similar cluster pair $(C_i, C_j)$ from the heap.
    \item If the similarity $\mathcal{H}(C_i, C_j)$ exceeds the current threshold $\tau$, we apply adaptive annealing:
    \begin{equation}
        \tau = \tau \times \alpha
    \end{equation}
    where $\alpha < 1$ is an annealing factor, and continue to the next iteration.
    
    \item Otherwise, we merge the clusters:
    \begin{equation}
        C_{\text{new}} = \text{Merge}(C_i, C_j)
    \end{equation}
    and add the new cluster to the hierarchical structure.
    
    \item For larger clusters, we apply contrastive refinement:
    \begin{equation}
        \text{if } |C_{\text{new}}| > m \text{ then apply refinement}
    \end{equation}
    This involves splitting the newly formed cluster:
    \begin{equation}
        \text{subclusters} = \text{Split}(C_{\text{new}}, \mathbf{A})
    \end{equation}
    and refining these subclusters using contrastive loss:
    \begin{equation}
        \text{RefineWithContrastiveLoss}(\text{subclusters}, \tau_{\text{contrast}})
    \end{equation}
Contrastive loss isn't implemented in the standard explicit form in traditional contrastive learning frameworks. Rather than computing an explicit loss value and performing gradient descent, the contrastive learning effect is achieved through the iterative centroid-based refinement process that reassigns points to their closest centroids, implicitly optimizing the same objective of minimizing intra-cluster distances and maximizing inter-cluster separation.    
    \item Finally, we update the heap with the new cluster's ANN neighbors.
\end{itemize}
In the final phase, we select representative prototypes for each leaf cluster:

\item For each leaf cluster $C_{\text{leaf}}$ in the hierarchy, we compute the centroid:
\begin{equation}
\boldsymbol{\mu} = \frac{1}{|C_{\text{leaf}}|} \sum_{x \in C_{\text{leaf}}} \mathbf{e}'_x
\end{equation}
\item We then select the top-3 utterances closest to the centroid as prototypes:
\begin{equation}
    \mathcal{R}[C_{\text{leaf}}] = \text{Top-3 } \arg\min_{x} \|\mathbf{e}'_x - \boldsymbol{\mu}\|_2
\end{equation}
These prototypes serve as representative examples of the intent represented by each leaf cluster. 
\end{itemize}

\subsection{Computational Complexity }
The algorithm achieves significant efficiency improvements over traditional hierarchical clustering methods. The use ANN for initial clustering reduces the complexity from $O(n^2)$ to $O(n \log n)$ and cluster merging from $O(n^3)$ to $o(k^2 \cdot \log(k)$).\\
The total complexity of the first phase for feature enhancements is: $O(n \cdot l^2 + n \cdot d \cdot h + n \cdot d) = O(n \cdot \max(l^2, d \cdot h))$\\
For ANN clustering : $O(n \cdot k \cdot d \cdot \log(n))$, which simplifies to $O(n \cdot \sqrt{n} \cdot d \cdot \log(n)) = O(n^{3/2} \cdot d \cdot \log(n))$ when $k = \sqrt{n}$. \par
Total complexity cluster merging is: $O(k^2 \cdot d + k^2 \cdot \log(k) + k^2 \cdot \log(M) + (k-1) \cdot (\log(M) + n + n^2 \cdot d + M \cdot d \cdot \log(k)))$
When $k = \sqrt{n}$, this simplifies to $O(n \cdot d + n \cdot \log(n) + n \cdot \log(M) + \sqrt{n} \cdot (n^2 \cdot d + M \cdot d \cdot \log(n)))$, which is dominated by the contrastive refinement term: $O(n^{2.5} \cdot d)$ in the worst case.\\
However, in practice, the contrastive refinement is only applied to large clusters, and the number of such refinements is typically much smaller than $k$, leading to a practical complexity closer to $O(n \cdot \log(n) + n \cdot d)$.

\section{Experimental Evaluation}
Our experimental evaluation systematically increases the dataset size and measures both stability and quality metrics at each increment. We visualize these trends using comprehensive plots showing the evolution of each metric.\par The food business-related 44112 questions are available in our created dataset with 255 business domain categories. Questions are embedded using the sentence-embedding method. For each run, we split the data into training and validation sets using an 80-20 split. The training set is incrementally sampled at predetermined sizes by the stratified Sampling approach, while the validation set remains constant to provide consistent evaluation.\par
We determine the optimal dataset size by identifying:
\begin{enumerate} 
\item When the cluster count stabilizes (derivative approaches zero)
\item When cluster movement falls below 5 percent.
\item When NMI and ARI exceed 0.85.
\item When quality metrics reach their optimal values.
\end{enumerate}
We conducted a comprehensive cluster stability analysis to determine the optimal dataset size for intent clustering. Figure~\ref{fig:fig2} illustrates the results of our stability metrics as we incrementally increased the dataset size. The hierarchical clustering algorithm implements a robust sampling procedure to ensure representative and balanced datasets for stability analysis. This sampling process is critical for accurately determining when the clustering structure stabilizes. 
\subsection{Stratified Sampling Approach}
The questions are sampling proportionally from each business category. The program intelligently adapts the sampling sizes based on data availability for Large Datasets (>30,000 samples) and uses larger step sizes: [10, 20, 40, 60, 80, 100, 120] samples per category. For Smaller Datasets, it dynamically calculates step sizes based on available data and creates approximately 6 evenly-spaced steps up to the maximum available data. Maximum balanced dataset possible: 36975 utterances for 44112 utterances and 35289 utterances are used as training samples, and  8823 used for validation samples.  This carefully designed sampling procedure ensures that the stability analysis accurately reflects how the clustering performance evolves with increasing dataset size. The stratified sampling distribution system uses dataset sizes: [2550, 5100, 10200, 15300, 20400, 25500, 30600]. 
\subsection{Result Analysis}
\begin{figure}[htbp]
  \centering
  \includegraphics[width=1.0\textwidth]{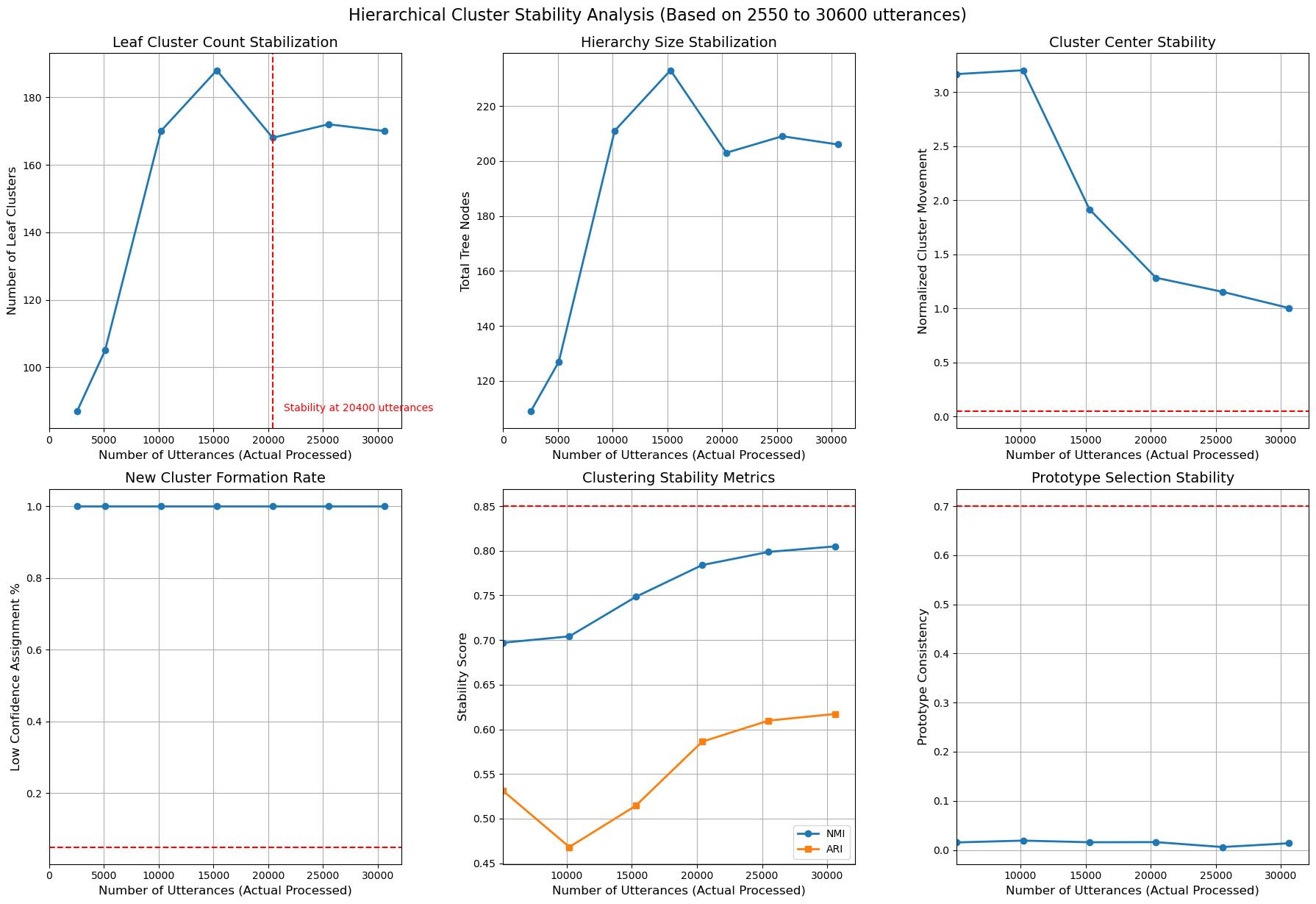} % Adjust width/path
  \caption{Stability and Performance of the Hierarchical Clustering}
  \label{fig:fig2}
\end{figure}

The cluster stabilization is measured by the various methods shown in Figure~\ref{fig:fig2}. The analysis of various stability metrics revealed that the clustering structure stabilized after approximately 20400 utterances.\par The leaf cluster stabilization only looks at the terminal nodes, and it is shown in Figure~\ref{fig:fig2}, which refers to the point of 20400 utterances with the derivative falling below our threshold of 0.001, indicating that additional data did not discover significant new intent categories. The hierarchy size stabilization graph refers to the point at which the total number of nodes in your tree structure stops growing significantly. The number of nodes in the hierarchy (including leaf and internal nodes) stabilized concurrently, demonstrating that the organizational relationships between intents had been adequately captured at the 10200-utterance point. The normalized movement of cluster centroids fell below 5\% after this threshold, indicating that the position of cluster centers in the embedding space had stabilized.\\ In the new cluster formation rate, the percentage of validation samples assigned consistently the same with low confidence. However, it will be good if it falls below a 5\% low-confidence score. Most utterances can be confidently assigned to existing clusters, suggesting good coverage of the intent space.\par
\begin{figure}[htbp]
  \centering
  \includegraphics[width=.90\textwidth]{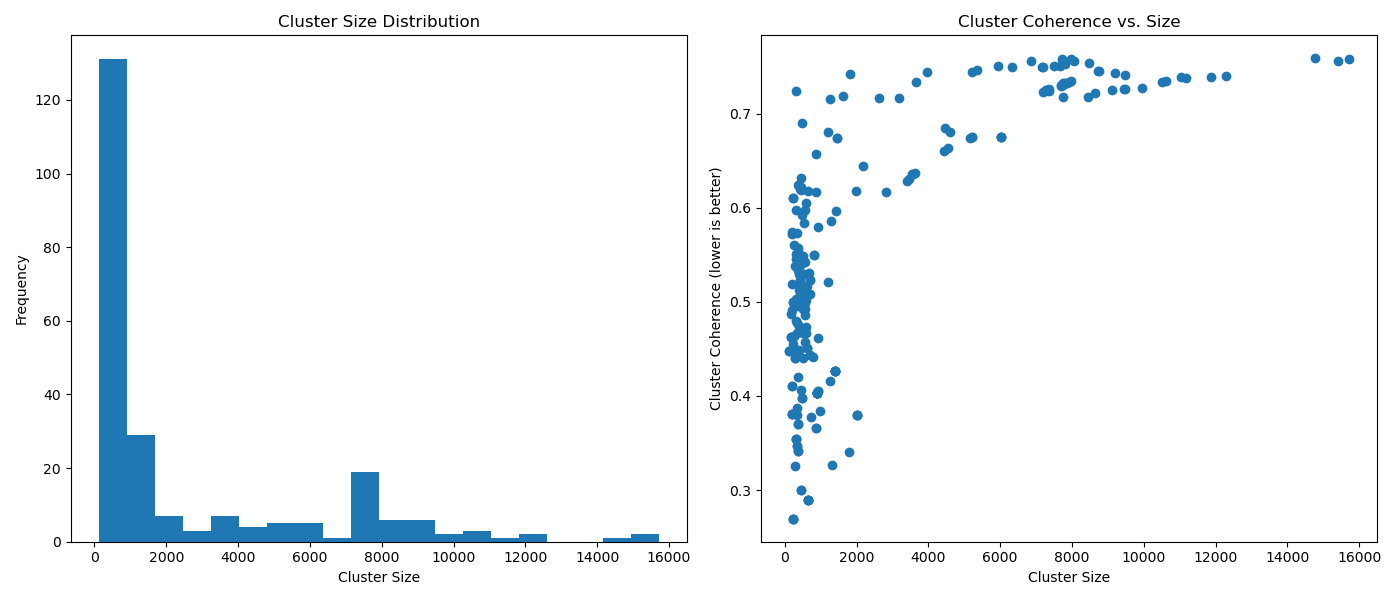} % Adjust width/path
  \caption{Clustering statistics}
  \label{fig:fig3}
\end{figure}

\begin{figure}[htbp]
  \centering
  \includegraphics[width=0.7\textwidth]{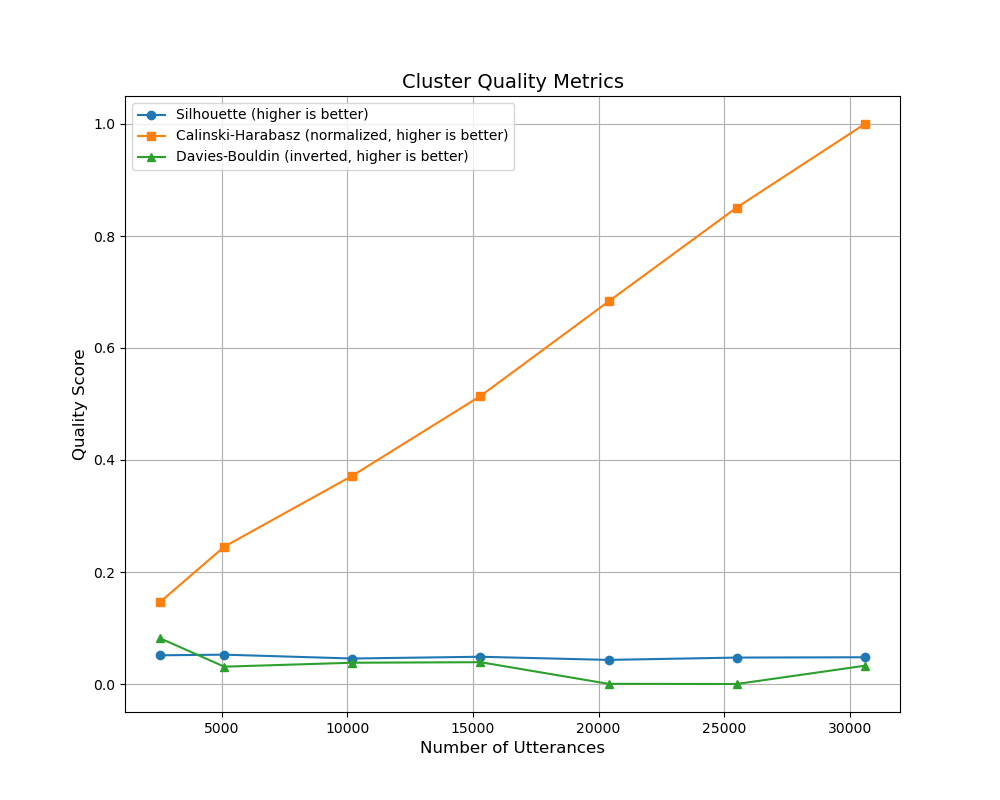} % Adjust width/path
  \caption{Qulaity metrics}
  \label{fig:fig4}
\end{figure}

 But the NMI stability score closure to .80 at this dataset size, indicating high consistency in cluster assignments between consecutive iterations, and similarly, the ARI stability score closure to 0.65 demonstrated high assignment consistency, confirming the stability of the clustering structure. The prototype consistency measures how stable the representative examples and label intent for each cluster remain when adding more data. Overall, our framework revealed that the use of a minimal domain intent with a hierarchical clustering structure tends to stabilize the domain intent knowledge structure.  \\
Cluster quality metrics are shown by the Figure~\ref{fig:fig4} in which indicates that the silhouette score is less than .1 value all the time and tends to 0, which means that intents are semantically overlapped. This result suggests the clusters have some overlap, which is expected for conversational intents. The Calinski-Harabasz (CH) Score is much higher, and the Davies-Bouldin (DB) Score close to 0 value both indicate that clusters are well-separated and well-defined.
Cluster distribution is shown in Figure~\ref{fig:fig3}, and cluster distribution refers to how questions are allocated across the different clusters in the hierarchy. As the dataset size increases, the number of clusters typically grows initially, then plateaus, suggesting that the ratio of questions per cluster tends to stabilize at an optimal dataset size. \\
Cluster coherence measures how similar questions within the same cluster are to each other, and typically, it is measured as the average distance of points to their cluster centroid. The Lower values indicate tighter and more coherent clusters. Smaller clusters tend to have better coherence, which means lower distance to the centroid, but when the dataset increases, coherence often decreases or becomes stable.\\
For our business domain intent clustering, we employed a human evaluation for the expert annotation process. The domain experts independently reviewed a sample of 500 intent clusters and their prototypes, as 3 questions as samples, rating the coherence and practical usefulness of each cluster. The accuracy for this was 92.5\%, and clusters received an average coherence score of 4.2/5.

Thus, it is concluded that our neural attention-based hierarchical clustering algorithm is more suitable for constructing a minimally supervised domain intent structure.

\section{Issues of Bias and Fairness in Intent Clustering}
Intent clustering systems can exhibit several critical biases that impact fairness:
\begin{enumerate}
    \item \textbf{Representational bias} occurs when specific intent categories are overrepresented in training data, leading to more refined clusters for majority intents while minority intents are grouped too broadly\cite{Johnson2021}.
    \item \textbf{Algorithmic bias} emerges when clustering methods favor certain linguistic patterns or expressions that are more common in specific demographic groups, resulting in inconsistent clustering quality across user populations\cite{Johnson2021}.
    \item\textbf{Embedding space distortion} \cite{Liang2023} happens when the vector representations used for clustering carry inherent biases from their training data, potentially placing semantically similar intents from underrepresented groups farther apart than they should be.
    \item \textbf{Granularity disparities} occur when the system creates detailed hierarchical structures for common domains while oversimplifying specialized or less frequently used intents, resulting in unequal service quality\cite{Hong2025}.
    \item \textbf{Prototype selection bias} affects system performance when the examples chosen to represent clusters primarily reflect mainstream expressions, potentially causing misclassification of valid but less common intent phrasings\cite{Liang2023}.
\end{enumerate}
These biases can collectively result in conversational systems that understand and respond more effectively to certain user groups while providing lower-quality service to others, reinforcing existing societal inequities in access to automated services.
\section{Conclusion and Future Work} 
This paper presents an efficient solution for building hierarchical domain intent structures in conversational recommendation systems using a neural attention-based approach. Our proposed algorithm successfully integrates concepts from DEC and NAM to create a robust hierarchical clustering method that adapts to evolving domain knowledge. The cluster quality and cluster statistics signified that the proposed algorithm is well-established in the domain intent structure. Through comprehensive stability analysis, we demonstrated that hierarchical intent structures can be effectively constructed with a minimal number of utterances (approximately 20400 utterances) when using stratified sampling strategies. The stability metrics—including cluster count stabilization, hierarchy size stabilization, and assignment consistency—provide strong evidence that our approach captures comprehensive domain intent coverage while maintaining coherent cluster structure.
The experimental results on 44K business domain questions confirm that our method significantly reduces the data requirements for building stable intent hierarchies. Our approach eliminates unnecessary data collection and labeling efforts by identifying the optimal, minimal dataset size while ensuring the system can accommodate new intents without requiring restructuring.\\ This efficiency directly contributes to more equitable AI solutions by lowering resource barriers for smaller businesses and organizations serving diverse communities. Our algorithm's significant reduction in data requirements (achieving stability with just 20,400 utterances) directly addresses societal equity by enabling underrepresented business domains to implement effective conversational recommendation systems despite limited resources. The optimization approach preserves domain-specific distinctions through our entropy-based merging strategy, while our attention mechanism ensures minority expressions receive appropriate recognition rather than being overshadowed by majority patterns. By lowering both computational and data collection barriers, our work democratizes access to an advanced conversational recommendation system for diverse business communities, aligning directly with the equitable solutions goal of creating more equitable AI solutions.\\
Our algorithm does not support learning in one domain, and after it can be transferred or adapted to new domains with minimal additional data. Future work will explore cross-domain adaptability, with a particular emphasis on how knowledge from well-represented domains can be efficiently transferred to underserved domains and communities with limited digital representation.

\begin{credits}
\subsubsection{\ackname} This (\textbf{Nourish}) research is funded by the \textbf{National Institution of Food and Agriculture U.S Department of Agriculture (USDA)}. The authors would like to acknowledge John Stephens for his valuable contributions to the methodological approach, whose insights significantly enhanced the development of the hierarchical domain intent tree presented in this work.

\end{credits}

\end{document}